\documentclass[10pt]{Final_transposition_llncs}

\usepackage{amsmath}
\usepackage{graphicx}
\usepackage{color}
\usepackage{url}
\usepackage{listings}
\usepackage{setspace}
\usepackage{subfigure}
\usepackage{wrapfig}
\usepackage{amssymb}
\usepackage{amsfonts}
\usepackage{rotating}
\usepackage{algorithm}
\usepackage{subfigure}
\usepackage{url}
\usepackage{pgf}
\usepackage{tikz}
\usepackage{listings}

\title{String comparison by transposition networks}
\author{Peter Krusche and Alexander Tiskin\\\small\email{[peter|tiskin]@dcs.warwick.ac.uk}}
\institute{Department of Computer Science, University of Warwick, Coventry CV4 7AL, UK}
\date{\today}

\lstdefinelanguage  {algorithm}
{keywords=[27]{begin,end,if,els e,elsif,then,while,for,repeat,do,input,output
,proc,return,let,with,where,true,false,exit,and,or,exists,no,such,next,find}, morecomment=[s]{\{}{\}} }
\lstnewenvironment {pseudocode}{\lstset{tabsize=4, language=algorithm,
texcl=true, mathescape=true, keywordstyle=\color{black}\bf,
basicstyle=\small\color{black}, commentstyle=\itshape }
\lstset{moredelim=[is][\underline]{_}{_}}
}{}

\begin{document}
\maketitle
\begin{abstract}
Computing string or sequence alignments is a classical me-thod of comparing
strings and has applications in many areas of computing, such as signal
processing and bioinformatics. Semi-local string alignment is a recent
generalisation of this method, in which the alignment of a given string and all
substrings of another string are computed simultaneously at no additional
asymptotic cost. In this paper, we show that there is a close connection
between semi-local string alignment and a certain class of  traditional
comparison networks known as transposition networks. The transposition
network approach can be used to represent different string comparison algorithms
in a unified form, and in some cases provides generalisations or improvements on
existing algorithms. This approach allows us to obtain new algorithms for sparse
semi-local string comparison and for comparison of highly similar and highly
dissimilar strings, as well as of run-length compressed strings. We conclude
that the transposition network method is a very general and flexible way of
understanding and improving different string comparison algorithms, as well as
their efficient implementation.
\end{abstract}

\section{Introduction}
In this paper we look at the classical problem of computing the cost of string
(or sequence) alignments, particularly the longest common subsequence 
and edit distance problems. Since this problem was originally 
proposed~\cite{Levenshtein:66}, a multitude of algorithms have been found to 
compute edit distances or equivalently longest common subsequences of two input
strings (see e.g.\ \cite{Hirschberg:97,Navarro:01} for an overview). 
An interesting extension to string alignment is semi-local string 
comparison. In this problem, we are interested in computing longest common subsequence lengths 
for one string and all substrings of the other string. Schmidt~\cite{Schmidt:98} proposed an 
algorithm for computing all longest paths in grid dags which was applied 
to string-substring longest common subsequence (LCS) computation by 
Alves et al.~\cite{Alves+:08}, who found an algorithm which runs in $O(n^2)$ time. 
Tiskin~\cite{Tiskin:05} developed
further understanding of the algorithm and its data structures, obtaining a subquadratic 
time algorithm for semi-local string comparison including string-substring and prefix-suffix 
LCS computation. Semi-local string comparison is useful as an intermediate step towards
fully-local string comparison, in which all pairs of substrings of the input strings
are compared. A straightforward application is e.g.\ computing the LCS efficiently 
in a sliding window (a slightly simpler version of this problem was studied 
in~\cite{Boasson+:99}).
Semi-local string comparison is also a useful tool for obtaining efficient parallel
algorithms for LCS computation~\cite{Apostolico:90,Alves+:06,KruscheTiskin:07}.
A summary of other algorithmic applications is given in~\cite{TiskinL:07:Applications}. 

In this paper, we develop a new interpretation of standard and semi-local LCS algorithms,  
based on a certain class of traditional comparison networks known as transposition 
networks. This approach allows us to obtain new algorithms for sparse semi-local string 
comparison and for comparison of highly similar and highly dissimilar strings, as well as 
semi-local comparison of run-length compressed strings. The remainder of this paper is 
structured as follows. We  introduce the necessary concepts of string comparison in
Section~\ref{sec:sequencecomparison}, and describe the transposition network method in 
Section~\ref{sec:transpositionnetworks}. We then show new algorithms for sparse semi-local 
string comparison in Section~\ref{sec:sparsegdags}, show how to compare run-length compressed 
strings semi-locally in Section~\ref{sec:rle}, and discuss comparing
highly similar or highly dissimilar strings in Section~\ref{sec:similiarity}.

\section{String comparison}\label{sec:sequencecomparison}
Let $x = x_1 x_2 \ldots x_m$ and $y = y_1 y_2 \ldots y_n$ be two strings over an
alphabet $\Sigma$ of size $\sigma$. We distinguish between consecutive
\textit{substrings} of a string $x$ which can be obtained by removing zero or
more characters from the
beginning and/or the end of $x$, and \textit{subsequences} which can be obtained by
deleting zero or more characters in any position.
The \textit{longest common subsequence} (LCS) of two strings is the longest
string that is a subsequence of both input strings, its length $p$ (the LLCS) is a
measure for the similarity of the two strings. 
Throughout this paper we will denote the set of integers 
$\{i, i+1, \ldots, j\}$ by $[i : j]$, and the set
$\{i+\frac{1}{2}, i+\frac{3}{2}, \ldots, j - \frac{1}{2}\}$ of odd
half-integers by $\langle i : j \rangle$. We will further mark odd
half-integer variables using a $\hat{\ }$ symbol.

\begin{figure}[t]
\begin{center}
  \includegraphics[width=0.8\textwidth]{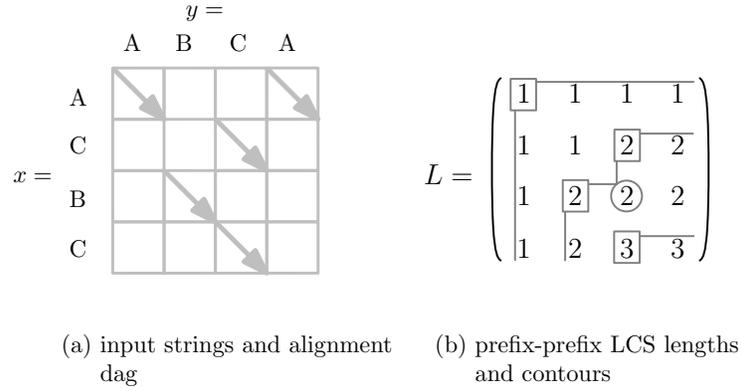}
  \caption{Example showing the alignment dag and the matrix $L$ of prefix-prefix LCS lengths}
  \label{fig:lcsexample}
\end{center}
\end{figure}
\begin{definition}
Let the \textit{alignment dag} be defined by a set of vertices $v_{i,j}$ with
$i \in [0:m]$ and $j \in [0:n]$ and edges as follows.
We have horizontal and vertical edges $v_{i, j-1} \rightarrow v_{i, j}$ and
$v_{i-1, j} \rightarrow v_{i, j}$ of weight~0. Further, we introduce diagonal edges 
$v_{i-1, j-1} \rightarrow v_{i, j}$ of weight~1 which are present only if 
$x_i = y_j$.\label{def:alignment dag}
\end{definition}
Longest common subsequences of a substring $x_i x_{i+1} \ldots x_{j}$ and $y$
correspond to longest paths in this graph from $v_{i-1,0}$ to $v_{j,m}$.
\begin{definition}
We define the \textit{extended alignment dag} as the infinite horizontal
extension of the alignment dag, having vertices $v_{i,j}$ as above, but allowing 
$j \in [-\infty:\infty]$, adding corresponding horizontal and vertical edges as 
above for all additional vertices, and further including diagonal edges $v_{i-1, j-1}
\rightarrow v_{i, j}$ of weight~1 for all $j < 1$ and $j > n$.
\label{def:extended alignment dag}
\end{definition}

For many applications, the LCS itself is of lesser interest than its length.
Looking at the LLCS for different substrings, including prefixes or suffixes of the 
input strings, exposes not only their global similarity, but also locations of 
high or low similarity. 
For example, the standard dynamic programming LCS algorithm
compares all prefixes of one string to all prefixes of the other
string~\cite{Wagner+:74} and stores their LCS lengths in the dynamic 
programming matrix $L(i,j) = \mathrm{LLCS}(x_1 \ldots x_i, y_1 \ldots y_j)$.
Semi-local string comparison~\cite{Tiskin:05} is an alternative to this standard
string alignment method. Solutions to the semi-local LCS problem are given
by a \textit{highest-score matrix} which we define as follows.

\begin{definition}
In a \textit{highest-score matrix} $A$, each entry $A(i,j)$ is defined as the LLCS
of $x$ and substring $y_i \ldots y_j$.
\end{definition}

\begin{figure}[t]
\begin{minipage}[c]{0.45\textwidth}
\begin{center}
\includegraphics[width=\textwidth,clip]{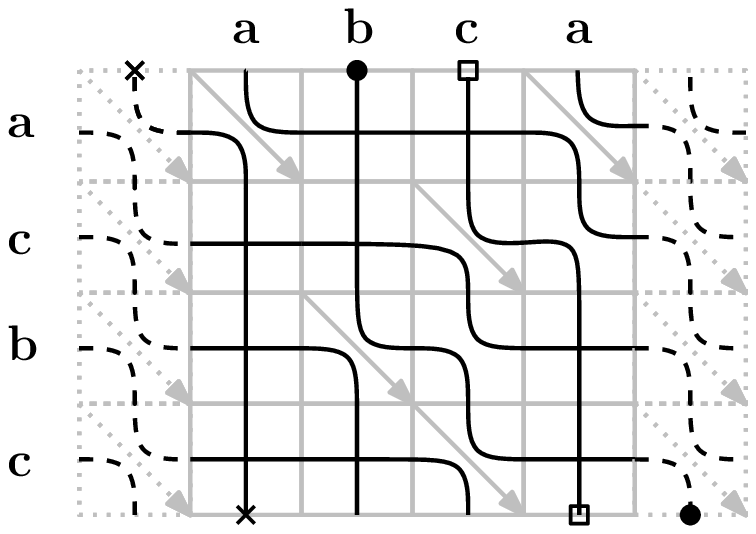}
\caption{\label{fig:seaweeds}Seaweed curves in the extended alignment dag}
\end{center}
\end{minipage}
\hfill
\begin{minipage}[c]{0.45\textwidth}
\begin{center}
\includegraphics[height=4cm,clip]{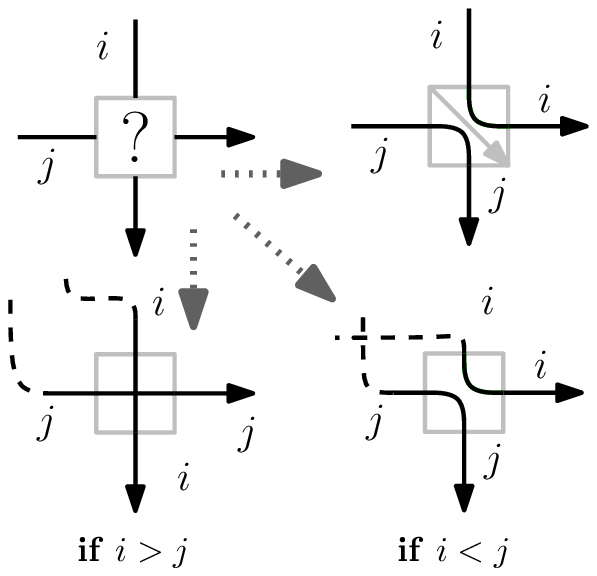}
\caption{\label{fig:seaweed_algorithm}Illustration of the seaweed algorithm}
\end{center}
\end{minipage}
\end{figure}
The definition of highest-score matrices can also be extended to include the LLCS
of all \textit{prefixes} $x_1 x_2 \ldots x_i$ and all \textit{suffixes} $y_j\ldots
y_n$, or the LLCS of all suffixes $x_i\ldots x_m$ and all prefixes $y_1\ldots
y_j$. 

Since the values of $A(i,j)$ for different $i$ and $j$ are strongly
correlated, it is possible to derive an implicit, space-efficient representation
of matrix $A(i,j)$. This implicit representation of a semi-local highest-score 
matrix consists of a
set of \textit{critical points}. 
\begin{definition}\label{def:critpoints}
The critical points of a highest-score matrix $A$ are defined as the set of odd
half-integer pairs $(\hat{\imath}, \hat{\jmath})$ such that
$A(\hat{\imath}+\frac{1}{2},\hat{\jmath}-\frac{1}{2}) + 1 = 
 A(\hat{\imath}-\frac{1}{2},\hat{\jmath}-\frac{1}{2}) =
 A(\hat{\imath}+\frac{1}{2},\hat{\jmath}+\frac{1}{2}) =   
 A(\hat{\imath}-\frac{1}{2},\hat{\jmath}+\frac{1}{2}) 
 $.
\end{definition}
Tiskin~\cite{Tiskin:05} showed that in order to represent a highest-score matrix 
for two strings of lengths $m$ and $n$, exactly $m+n$ such critical points are necessary. 
Note that infinitely many critical points exist in the extended alignment dag. 
However, due to the structure of
the extended alignment dag, only a \textit{core} of $m+n$ critical points need to be
stored. Each of the remaining \textit{off-core} critical points can be computed in 
constant time. 
\begin{theorem}\label{thm:implicitrep}
The highest-score matrix $A$ can be represented implicitly using only  $O(m+n)$ space by its core critical points. 
We have: $A(i,j) = j - i - | \{(\hat{\imath}, \hat{\jmath})\ :\ 
(\hat{\imath}, \hat{\jmath})\text{ is a critical point, } i < \hat{\imath} \text{ and } \hat{\jmath} < j\}|$.
\end{theorem}
\begin{proof}
See~\cite{Tiskin:05}. 
\end{proof}
Theorem~\ref{thm:implicitrep} is a direct
consequence of the Monge properties~\cite{Burkard+:96} of highest-score matrices. 
This theorem is particularly useful as it was also shown possible to combine two 
highest-score matrices in subquadratic time using their implicit 
representation~\cite{Tiskin:05} in order to obtain the
highest-score matrix corresponding to comparing one string and a concatenation of two 
other strings.
In~\cite{Alves+:06,Tiskin:05:ParCo,KruscheTiskin:07}, these methods were applied to obtaining
efficient parallel algorithms for the LCS problem.

The set of critical points can be obtained using the \textit{seaweed algorithm} 
(by Alves et al.~\cite{Alves+:08}, based on Schmidt~\cite{Schmidt:98}, adapted by 
Tiskin~\cite{Tiskin:05}) which computes critical points incrementally
for all prefixes of the input strings. This dynamic programming procedure is 
graphically illustrated by tracing \emph{seaweed curves} that start at odd half-integer 
positions between two adjacent vertices $v_{0, \hat{\imath} - \frac{1}{2}}$ and
$v_{0,\hat{\imath} + \frac{1}{2}}$ in the top row of the extended alignment dag,
and end between two adjacent vertices $v_{m, \hat{\jmath} - \frac{1}{2}}$ and
$v_{m,\hat{\jmath} + \frac{1}{2}}$ in the bottom row (see Figure~\ref{fig:seaweeds}).
Each critical point is computed as the pair of horizontal start and end coordinates 
of such a seaweed curve.

\begin{definition}
Given the sequence $\{(\hat{\imath}, \hat{\jmath}_k)\ :\ k\in [1:m]\}$
where $(\hat{\imath}, \hat{\jmath}_k)$ is a critical point in the highest-score
matrix of $x_1\ldots x_k$ and $y$, a seaweed curve is obtained by connecting
the sequence of points $\{(\hat{\imath},\hat{\jmath}_1), (\hat{\jmath}_1,\hat{\jmath}_2),
\ldots,$ $(\hat{\jmath}_{m-1},\hat{\jmath}_m)\}$.
\end{definition}

When drawing the (extended) alignment dag in the plane, its horizontal and vertical 
edges partition the plane into rectangular cells which, depending on the input
strings, may contain a diagonal edge or not. 

\begin{definition}
For every pair of characters $x_i$ and $y_j$ we define a corresponding
\textit{cell} $(i-\frac{1}{2}, j-\frac{1}{2})$. Cells corresponding to a
matching pair of characters will be called \textit{match cells}, cells 
corresponding to mismatching characters or to cells only present in the 
extended alignment dag will be called \textit{mismatch cells}.
\end{definition}

Two seaweed curves enter every cell in the extended alignment dag, one at
the left and one at the top. The seaweed curves proceed through the cell
either downwards or rightwards. In the cell, the directions of these curves are
interchanged either if there is a match $x_{k} = y_{l}$, or if the same
pair of seaweed curves have already crossed. Otherwise, their directions remain
unchanged and the curves cross. The seaweed algorithm is illustrated in 
Figure~\ref{fig:seaweed_algorithm}.

More efficient special case algorithms for the LCS problem can be obtained when parameterizing
either by the number $r$ of match cells, by the length $p$ of the LCS, or by the edit distance.
Previously, high-similarity string comparison has been considered in
\cite{Hirschberg:77,Nakatsu+:82,Ukkonen:85,Myers:86,Kumar_Rangan:87,Wu+:90,Rick:95,Rick:00};
all these papers give LCS algorithms for highly similar strings, running in time
$O(ne)$, where $e$ is either the edit distance between the strings (as in
\cite{Ukkonen:85}), or a different closely related similarity measure.
High-dissimilarity string comparison has been considered in
\cite{Hirschberg:77,Rick:95,Rick:00}; the best running time for LCS on highly
dissimilar strings is $O(np + n\log n)$.
A good survey of parameterized string comparison algorithms is given by~\cite{Hirschberg:97}.

The basis of parameterized LCS computation for dissimilar strings is to determine 
the LCS of two strings as a longest \textit{chain} of match cells $(i_1, j_1), (i_2, j_2),
\ldots,$ $(i_p, j_p)$ with $i_1 < i_2 < \ldots < i_p$ and $j_1 < j_2 < \ldots <
j_p$. We define a partial order on the set of match cells by $(i_1,j_1) \prec
(i_2,j_2)$ iff. $i_1 < i_2$ and $j_1 < j_2$; further, we say that $(i_1,j_1)$ is
dominated by $(i_2,j_2)$.  Due to Dilworth's lemma~\cite{Dilworth:50}, the
minimum number of antichains (sets of pairwise incomparable elements) necessary to
cover a partially ordered set is equal to the length of the longest chain.
Therefore, the LCS of two strings can be obtained by computing a minimal antichain
decomposition of the set of matches under the $\prec$ ordering. 
Consider chains ending at a match $(i,j)$. If any longest such chain has
length $k$, then this match is said to have \textit{rank} $k$. If match $(i,j)$ has rank
$k$ and for all other matches $(i',j')$ of rank $k$ either $i' \geq i$ and $j' < j$ or
$j' \geq j$ and $i' < i$, then match $(i,j)$ is called ($k$-)dominant.

The set of all dominant matches completely specifies the table of prefix-prefix LCS lengths 
$L(i,j) = LLCS(x_1\ldots x_i,y_1\ldots y_j)$.
Let the contours of $L$ be formed by the rows and columns of cells through which the values 
of $L$ increase by one. A cell $(\hat{\imath}, \hat{\jmath})$ belongs to a contour in $L$ 
if $L(\hat{\imath} + \frac{1}{2}, \hat{\jmath} + \frac{1}{2}) > L(\hat{\imath} - \frac{1}{2}, \hat{\jmath} +
\frac{1}{2})$, $L(\hat{\imath} + \frac{1}{2}, \hat{\jmath} + \frac{1}{2}) > L(\hat{\imath} + \frac{1}{2},
\hat{\jmath} - \frac{1}{2})$, or $L(\hat{\imath} + \frac{1}{2}, \hat{\jmath} + \frac{1}{2}) > 
L(\hat{\imath} - \frac{1}{2}, \hat{\jmath} - \frac{1}{2})$. Figure~\ref{fig:lcsexample}~(b)
shows an example.
All match cells belonging to the same contour form an antichain in a minimal antichain 
decomposition, and each contour is specified completely by the dominant
matches on it\footnote{In~\cite{Rick:95,Rick:00}, these contours are called forward contours.}.

Since parameterized algorithms process the input match-by-match instead of
computing the entire prefix-prefix LCS score matrix, it is necessary to
pre-process the input strings to obtain lists of match cells. Different 
approaches exist for this, depending on the assumptions that can be made about the
alphabet. Generally, it is necessary to allow less-than/greater-than comparisons
in addition to testing for equality (otherwise, $\Omega(mn)$ was shown to be a lower
bound~\cite{Aho+:76_JACM}). Based on this assumption, we can obtain a set of
match lists which give for every character $c$ in $x$ the positions $i$ where $y_i = c$ in
$O(n\log n)$ time. These lists usually allow queries for increasing or decreasing
values of $i$ and are called occurrence lists or match lists accordingly.  
The lists are obtained by determining the inverse sorting permutation for $y$ 
(i.e.\ a permutation that transforms a sequence which contains all characters from $y$ 
in sorted order into $y$). For every character $c$ in $x$, we can find the head of a list of match
positions in time $O(\log n)$ by binary search. For small alphabets, it is
possible to pre-process the input in time $O(n\log \sigma)$ to obtain a similar
representation (see \cite{Hirschberg:77,Apostolico:97} for discussion). We will
denote the result of this preprocessing as follows.
\begin{definition}
The functions $\mu_i : \mathbb{N} \rightarrow [1:n] \cup {\infty} $ for $i \in [1:m]$
specify the match positions. We have:
\begin{itemize}
  \item $\mu_i(k) = j, j \neq \infty \Rightarrow x_i = y_j$,
  \item $\mu_i(k) < \mu_i(k+1)$ or $\mu_i(k) = \mu_i(k+1) = \infty$ for all $k \in [1:n-1]$.
\end{itemize}
\end{definition}
This notation allows storing the match lists using $O(m+n)$ space. 
We can obtain these functions for arbitrary ordered alphabets in time $O(n\log n)$
by sorting one of the input strings and then using binary search to create the match
lists.
For small alphabets of size $\sigma < n$, the sorting permutation can be determined in 
time $O(n\log \sigma)$ by counting character frequencies for all characters
contained in $y$. After this pre-processing step, we can determine $\mu_i(k)$ in $O(1)$ time using $O(m+n)$ storage.

\section{The transposition network method}\label{sec:transpositionnetworks}
Comparison networks (see e.g.~\cite{Cormen+01}) are a traditional
 method for studying oblivious algorithms for sorting
sequences of numbers. A \textit{comparison network} has $n$ inputs and $n$
outputs, which are connected by an arbitrary number of \textit{comparators}.
A comparator has two inputs and two outputs. It compares the input values and
returns the larger value at a prescribed output, and the smaller value at the
other output. We will draw comparison networks as $n$ wires, where pairs of wires
may be connected by comparators that operate on the values passing through the
wires. Comparators are usually grouped into a
sequence of $k$ \textit{stages}, where each wire is connected to at most one
comparator in a single stage. A comparison network is called
\textit{transposition network} if all comparators only connect adjacent wires.

\begin{figure}[t]
\begin{minipage}[t]{0.45\textwidth}
\begin{center}
\includegraphics[height=4cm,clip]{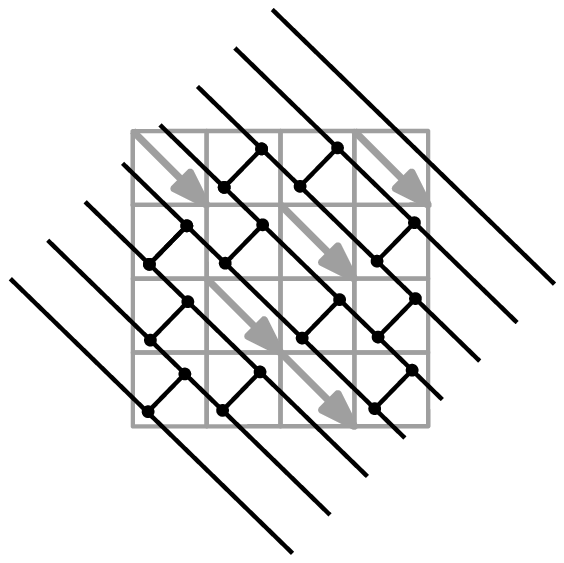}
\caption{\label{fig:mergingnetwork}Comparison network of an alignment dag}
\end{center}
\end{minipage}
\hfill
\begin{minipage}[t]{0.42\textwidth}
\begin{center}
\includegraphics[width=\textwidth,clip]{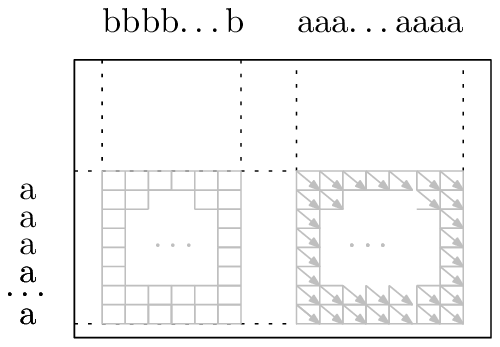}
\caption{Alignment dag for run-length compressed strings}\label{fig:rle_gdag}
\end{center}
\end{minipage}
\end{figure}

Transposition networks allow for another interpretation
of the seaweed algorithm. As shown in Figure~\ref{fig:mergingnetwork},
every mismatch cell behaves like a comparator on the
starting points of the seaweeds that enter the cell from the left and the top.
The larger value is returned on the right output,
and the smaller value is returned on the bottom output.
For a match cell, the input values are not compared but just translated top to
right and left to bottom.
Therefore, we can define a transposition network for every problem instance
as follows.

\begin{definition}
The network $\mathrm{LCSNET}(x,y)$ has $m+n$ diagonal wires.
Every mismatch cell
$(\hat{\imath}, \hat{\jmath})$ corresponds to a comparator in stage
$\hat{\imath}+\hat{\jmath}$ connecting wires $m-\hat{\imath}+\hat{\jmath}$ and
$m-\hat{\imath}+\hat{\jmath}+1$ (see Figure~\ref{fig:mergingnetwork}). Match
cells do not contain comparators.
\end{definition}

As comparators in the network correspond to
cells in the alignment dag, we choose the convention of drawing the network
wires top left to bottom right. Values moving through a cell or comparator can
therefore move either down or to the right.

The network $\mathrm{LCSNET}(x,y)$ realizes the seaweed algorithm. The inputs are
originally in inversely (in relation to the direction of the comparators) sorted
order and trace the seaweed curves on their paths through the transposition
network. Note that the direction of the comparators can be determined arbitrarily,
as long as it is opposite to the sorting of the input sequence. Another degree
of freedom when defining transposition networks lies in the behaviour of
comparators for equal inputs. Even though this does not affect the network
output, changing the convention of swapping or not swapping equal values can
simplify specification of non-oblivious algorithms for computing the output
values.

In order to solve the global or semi-local LCS problem for strings $x$ and $y$
using the transposition network method, we have to define appropriate input
values for $\mathrm{LCSNET}(x,y)$. In order to obtain the full set of critical
points, the inputs are set to the seaweed starting points:
input $\hat{\jmath}+m+\frac{1}{2}$ is initialized with $\hat{\jmath}$, $\hat{\jmath} \in
\langle-m:n\rangle$. Let the vector $O$ denote the output of the network.
If all comparators return the larger input on the bottom
output, and the smaller input on the right output, the pairs
$(O(\hat{\jmath}+m+\frac{1}{2}), \hat{\jmath}+m)$ with
$\hat{\jmath} \in \langle -m:n \rangle$
correspond to the core critical points of the corresponding highest-score matrix.
Since there are $O(m n)$ comparators in the transposition network, the
resulting algorithm runs in time $O(m n)$.

Using the transposition network method, we can see the connection between semi-local
string comparison and existing LCS algorithms is the fact that both approaches
compute LCS scores incrementally for prefixes of the input strings: the standard
LCS dynamic programming approach
computes LCS lengths, and the seaweed algorithm computes implicit highest score
matrices for all prefixes of the input strings. When looking at this
relationship in more detail, it becomes clear that standard LCS algorithms can
be obtained by the transposition network method using input values of only zero
or one. A first direct consequence are bit-parallel LCS
algorithms~\cite{Crochemore+:01,Crochemore+:03}, which can be obtained by
computing the output of the transposition network cell-column by cell-column
using bit-vector boolean operations and bit-vector addition. In the remainder
of this paper we will show further examples where existing algorithms for
comparing two strings globally can be derived from transposition networks,
and discuss generalizing them to semi-local string comparison.

\section{Sparse semi-local string comparison}\label{sec:sparsegdags}
We now consider \emph{sparse string comparison}, i.e.\ string comparison
parameterized by the number of matches $r$ in the alignment dag.
Hunt and Szymanski~\cite{Hunt:77a} proposed an algorithm for sparse string
comparison that computes the LCS of two input strings in  $O((r+n) \log n)$ time.
An extreme case of this is the comparison of permutation
strings of length $n$ over the alphabet $\Sigma = [1:n]$. In this case, only $n$ match 
cells exist. Tiskin~\cite{Tiskin:06Permutations} gave an $O(n^{1.5})$ algorithm for semi-local
comparison of permutation strings.

Since in sparse string comparison the alignment dag contains few matches, large
rectangular areas of the transposition network have full sets of comparators. 
These areas will be denoted as follows.
\begin{definition}
Let network $\mathrm{DIAMOND}(m,n)$ be defined as an $\mathrm{LCSNET}$ network
which corresponds to a problem instance with no matches. It therefore contains a full
set of $m\cdot n$ comparators.
\end{definition}
We now give a more general sparse semi-local string comparison algorithm 
parameterized by the number of matches. We will first show a non-oblivious algorithm to compute the output of
$\mathrm{DIAMOND}$ networks efficiently, and then propose a technique for evaluating a $\mathrm{LCSNET}$ network
by partitioning it into smaller $\mathrm{DIAMOND}$ networks.

Consider an $m' \times n'$ rectangular area in the alignment dag with only mismatch
cells, and the corresponding $\mathrm{DIAMOND}(m', n')$ network.
Such an area occurs whenever two substrings over disjoint character sets are
compared. The network consists of a full set of $m'\times n'$
comparators and $m'+n'$ wires.\footnote{Note that the opposite case of a
rectangular area in the alignment dag which only contains matches is trivially
solved in linear time as it corresponds to a transposition network
without comparators.}
If the first $m'$ and the following $n'$ wires are initialized with
two pre-sorted sequences of numbers, this network works as a merging
network~\cite{Munter:93}. The problem of merging pre-sorted sequences can be
solved non-obliviously in time $O(m'+n')$. However, as the inputs to the
$\mathrm{DIAMOND}$ network are not necessarily pre-sorted, this is not
sufficient.
\begin{theorem}\label{thm:diamondnet}
It is possible to compute the outputs of the $\mathrm{DIAMOND}$ network
non-obliviously in time $O((m'+n')\log (m'+n'))$ if the inputs are in
arbitrary order. Additionally, if the sorting permutation of the inputs is
known (but the inputs are still in arbitrary order), the problem
can be solved in $O(m'+n')$  time, as the factor of $\log (m'+n')$
only comes from the initial sorting step.
\end{theorem}
\begin{proof}
To non-obliviously compute the
output of $\mathrm{DIAMOND}(m',n')$,
consider the path that the largest input takes through the network. If the largest input
enters the network on wire $j$, all comparators it passes will return it as the
larger element, which means that it will reach the leftmost output possible.
We then proceed through the remaining inputs in descending order, determining for every
input the leftmost output it can reach, considering that some outputs have
already been occupied by larger values.
Any current value that enters the comparison network on a wire $j$ that is less than
$m'$ wires ahead of the first free output will be translated to the first (leftmost)
available output.
If the current value enters the network more than $m'$ wires to the right of
the first available output, it can only pass through $m'$ comparators and will therefore
reach output $j-m'$.
The free outputs are indicated by a Boolean array $K$, where occupied outputs are marked with
a value of true. Since we proceed through the input values in descending order,
this yields the same output as direct evaluation of the transposition network. 
The entire algorithm is shown in Algorithm~\ref{alg:diamondnet_solve_2}.
$\Box$
\end{proof}
\begin{algorithm}[t]
 \caption{Computing the output of $\mathrm{DIAMOND}(m',n')$}\label{alg:diamondnet_solve_2}
\begin{pseudocode}
input: $I[1], \ldots, I[m'+n']$
output: $O[1], \ldots, O[m'+n']$
let $I[L[1]] > I[L[2]] > \ldots > I[L[m'+n']]$ { $L \text{ is the sorting permutation}$ }
for $(j \in [1:m'+n'])$ do $K[j] \leftarrow $ false { $K \text{ contains the non-free outputs}$ }
$\beta \leftarrow 1$ { $\beta \text{ points to the leftmost free output}$ }
for $k = 1, 2, \ldots m'+n'$ { $I[L[k]] \text{ is the next largest element}$ }
 if $L[k] < \beta + m'$ then { $\text{Maximum reaches leftmost free output}$ }
  $O[\beta] \leftarrow I[L[k]]$  { $\text{Translate input value to output}$ }
  $K[\beta] = $ true { $\text{Mark output as occupied}$ }
  while $(K[\beta]$) do $\beta \leftarrow \beta + 1$
 else { $\text{Maximum goes to leftmost output it can reach}$ }
  $O[L[k] - m'] \leftarrow I[L[k]]$ { $\text{Translate input value to output}$ }
  $K[L[k] - m'] = $ true { $\text{Mark output as occupied}$ }
 end if
end while
\end{pseudocode}
\end{algorithm}

Using Algorithm~\ref{alg:diamondnet_solve_2}, we obtain an improved
algorithm for sparse semi-local comparison. For simplicity assume that
both strings are of length $n$ and (w.l.o.g.) that $n$ is a power of~2.
\begin{theorem}
After pre-processing the input strings for obtaining match lists, the problem of semi-local string comparison can be
solved in $O(n\sqrt{r})$ time.
\end{theorem}
\begin{proof}
We first find the sorting permutations of the input strings.
This is possible in time $O((m+n)\log\min(\sigma, \max(m,n)))$, similar to 
obtaining $\mu_i$ in Section~\ref{sec:sequencecomparison}.

After this pre-processing, we partition the alignment dag into blocks using 
a recursive quadtree scheme. Consider processing such a block of size $w\times w$. Let 
this block correspond to comparing substrings $x_k\ldots x_{k+w-1}$ and $y_l\ldots y_{l+w-1}$.
As an input for each such block, we have the sorting permutations of the 
two corresponding substrings, the input values for the transposition network 
corresponding to the block, and also the sorting permutation for these input values.
For each block, we obtain the output values of its transposition network and their
sorting permutation as follows.

For a $w\times w$ block, we can count the number of matches in it in time
$O(w)$ by linear search in the sorting permutations of the corresponding
substrings.  Whenever we find a block that does not contain any matches, we stop
partitioning and use Algorithm~\ref{alg:diamondnet_solve_2} to compute the
outputs of the corresponding comparison network. Otherwise, we continue to
partition until we obtain a $1\times 1$ block that only consists of a single match.

A $1\times 1$ leaf block consisting of a single match can be processed
trivially in constant time.
Due to Theorem~\ref{thm:diamondnet}, we can compute the outputs for a
$w\times w$ mismatch block in $O(w)$ time when the sorting permutation
is known for the inputs. The sorting permutation for the 
root block of the quadtree is known, 
since the root of the quadtree corresponds to the full alignment dag, and the
inputs to its transposition network form a sequence sorted in reverse.
For all other blocks, we keep track of the
sorting permutation of both its input and output elements. 
For every output we can trace the input it came from before executing 
Algorithm~\ref{alg:diamondnet_solve_2} and therefore know the permutation 
that was performed by the transposition network within the block. 
Knowing this permutation and the sorting permutation
of the inputs allows to establish the sorting permutation of the outputs
in time $O(w)$.

To summarize, given the input values and their sorting permutation for every
leaf block of the quadtree recursion, we can compute the output values
and their sorting permutation in time $O(w)$.
All non-leaf blocks are partitioned into four
sub-blocks of size $w/2 \times w/2$. The inputs and their sorting permutation are
split and used to recursively process the sub-blocks. We can then establish the
sorting permutation of the outputs for the entire block in linear time by merging. To
compute the outputs of any intermediate block we therefore need time $O(w)$ in
addition to the time necessary for recursively processing the sub-blocks.

Consider the top $\log_4 r$ levels of the quadtree. In each subsequent level,
the number of blocks increases by at most a factor of four, and the block size
decreases by a factor of two. Therefore, this part of the quadtree is dominated
by level $\log_4 r$ which contains at most $r$ blocks, each of size $n/\sqrt{r}$.
The total work required on this part of the tree is therefore
$O(r \cdot n/\sqrt{r}) = O(n\sqrt{r})$.

The remaining levels of the quadtree can each have at most $r$ blocks
that still contain matches. The block size in each level still decreases
by a factor of two. Therefore, this part of the quadtree is also dominated
by level $\log_4 r$ and requires the same asymptotic amount of work.
The overall time for the algorithm is therefore bounded by
$\sum_{j = 0}^{\log_4 r} O(n/2^j \cdot 4^j) +
\sum_{j = \log_4 r + 1}^{\log_4 n} O(n/2^j \cdot r)
 = O(n\sqrt{r})\text{.} $
The resulting algorithm has running time $O(n\sqrt{r})$, and thus
provides a smooth transition between the dense case ($r = n^2$,
running time $O(n^2)$) and the permutation case ($r = n$, running time $O(n^{1.5})$). $\Box$
\end{proof}

\section{Semi-local comparison of run-length compressed strings}\label{sec:rle}
Another straightforward application of Algorithm~\ref{alg:diamondnet_solve_2} is comparing
run-length compressed strings~\cite{Apostolico+:99}. In this compression method, a run of
repeating characters is encoded by a single character together with the number
of repetitions. A run-length encoded string  $X = X_1 X_2 X_3\ldots
X_{\overline{m}}$ consists of $\overline{m}$ character runs $X_j$ of lengths
$|X_j|$. The length of the full string is therefore $m = \sum_{j = 1\ldots
\overline{m}} |X_j|$. When constructing the alignment dag for comparing two
run-length compressed strings $X= X_1 X_2 X_3\ldots X_{\overline{m}}$ and $Y=
Y_1 Y_2 Y_3\ldots Y_{\overline{n}}$,
rectangular areas without matches occur when character runs in $X$ and $Y$ mismatch.
Analogously, large rectangular areas with containing only match cells occur if the characters do match 
(see Figure~\ref{fig:rle_gdag}). Using the comparison network method and
Algorithm~\ref{alg:diamondnet_solve_2}, these rectangular areas can be
processed in cost proportional to their perimeter. Given two input strings with
uncompressed lengths $m$ and $n$, and compressed lengths $\overline{m}$ and
$\overline{n}$, this method results in an algorithm for semi-local comparison
which has cost $\sum_{i \in [1:\overline{m}], j\in [1:\overline{n}]} O(|X_i|+|Y_j|) =
O(\overline{m}n + m\overline{n})$.
This is as good as the result from~\cite{Bunke_Csirik:93}, additionally solving
the more general problem of semi-local string comparison of run-length compressed strings.

\section{High similarity and dissimilarity string comparison}\label{sec:similiarity}
In Section~\ref{sec:sparsegdags} we described an efficient algorithm for semi-local
string comparison, parameterized by the overall number of matches. We now
describe an application of the transposition network method to designing
algorithms that are parameterized by the LCS length $p$ of the input strings
or their LCS distance $k = n-p$. Such parametrization provides efficient algorithms
when the corresponding parameter is low, i.e.\ when the strings are 
highly dissimilar or highly similar.

In~\cite{Hunt:77a}, matches are processed row by row to establish
which antichain they belong to. Apostolico and Guerra improved this algorithm by
avoiding the need to consider non-dominant matches~\cite{Apostolico:87} (see Section~\ref{sec:sequencecomparison}), 
and changing the order in which the match cells are processed. This allows to obtain 
an algorithm that is parameterized by the length of the LCS. Further, there have been 
various extensions to this approach, which improve the running time by either using 
different data structures~\cite{Eppstein+:92} or narrowing the area in which to search for
dominant matches hence giving algorithms which are efficient both when
the LCS of the two strings is long or short~\cite{Rick:95}. In this paper, 
we will show how the transposition network method can be used to match these
algorithms for global LCS computation. For semi-local alignment, we achieve
a running time of $O(np)$, which is efficient for dissimilar strings. 

We will now show the connection between the antichain decomposition of the
set of match cells and the transposition network method.
Consider an $\mathrm{LCSNET}$ network with the following input values:
The first $m$ wires (i.e.\ the inputs on left hand side of the alignment dag)
are initialized with ones, and the following $n$ wires (i.e.\ the inputs at the
top of the alignment dag) are filled with zeros. On all comparators, smaller
values are returned at the bottom output. We will refer to this specific
transposition network setup as \textit{$\mathit{LCSNET}(x,y)$ with 0/1 inputs}. 
Using only zeros and ones as inputs to $\mathrm{LCSNET}(x,y)$ corresponds to 
tracing seaweeds anonymously, only distinguishing between those seaweeds that start
at the top and those seaweeds that start at the left.
The 0-1 transposition network approach allows to understand previous results
for parameterized LCS computation in terms of transposition networks, and helps to extend
some of these to semi-local string comparison.
\begin{corollary}\label{cor:0-1 net output}
In $\mathrm{LCSNET}(x,y)$ with 0/1 inputs as described above, let $p$ be the number of 
ones reaching output wires below $m+1$ (i.e.\ the bottom of the alignment dag). 
This number is equal to the number of zeros reaching an output wire above $m$ (i.e.\ the right side
of the alignment dag), and $LLCS(x,y) = p$.
\end{corollary}
\begin{proof}
From Theorem~\ref{thm:implicitrep}, we know that $LLCS(x,y) = n - d$, where $d$ is the number
of seaweeds that start at the top and end at the bottom of the alignment dag. 
The number of zeros ending up at the bottom is therefore equal to~$d$, and the number of
ones ending up at the bottom is equal to $n-d = LLCS(x,y)$. Since the transposition
network outputs a permutation of the input, and since we have $n$ input zeros, $n-d$~zeros 
must end up at the right. $\Box$
\end{proof}

We will now look at the behaviour of $\mathrm{LCSNET}(x,y)$ with 0/1 inputs in more detail.
In order to be able to trace paths of individual values, we must specify the behaviour
of the comparators for equal input values (note that changing this specification does not 
change the output of $\mathrm{LCSNET}(x,y)$). 
Assume that comparators in $\mathrm{LCSNET}(x,y)$ swap their input values if these
are equal.
If the alignment dag contains only mismatch cells and therefore a full set of comparators, 
all ones move from the left to the right, and all zeros move from the top to the bottom.
When introducing a match cell and hence removing a comparator, the zero that enters the 
match cell at the top is translated to the right, and the value of one entering the match cell at
the left is translated to the bottom. We trace these two values
further: as identical values are swapped by convention, both the one (and equally the zero) 
will not change direction of movement and be passed on vertically (horizontally in case of 
the zero) through all comparators. We will refer to ones which move downwards and to zeros
which move to the right as \textit{stray}.   
Stray values only change direction again when they either encounter a match cell or another
stray value. If two stray values enter the same cell, they leave this cell in the original 
directions, the one moving rightwards, and the zero moving downwards. This happens 
independently of whether this cell contains a match: in a match cell, no 
comparison is performed, the stray zero is returned at the bottom and the stray one is 
returned at the right. In a mismatch cell, the zero is also returned at the bottom 
since it is the smaller value. Therefore, two stray values always return to their original direction
of movement when meeting in the same cell. Another observation is that any cell which
has exactly one stray input value must have equal inputs. If such a cell is a match
cell, the stray input value returns to its original direction of movement, and
the other input becomes stray. If the cell does not contain a match, the inputs
are exchanged by convention, and the stray value remains stray. To summarize,
stray values caused by a match cell will start a row (stray zeros) or column (stray
ones) of cells which output stray values. This row or column only ends when meeting another column
or row of cells which output stray values. 

\begin{figure}[t]
\begin{minipage}[t]{0.45\textwidth}
\begin{center} 
 \includegraphics[height=4cm]{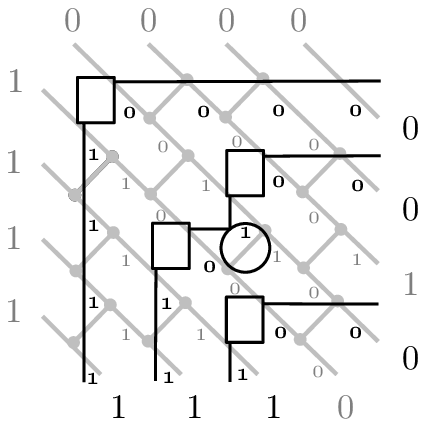}
\end{center}
 \caption{$\mathrm{LCSNET}(x,y)$ with 0/1 inputs}
 \label{fig:zeroone_example}
\end{minipage} 
\hspace{5mm}
\begin{minipage}[t]{0.43\textwidth}
\begin{center}
  \includegraphics[height=4cm]{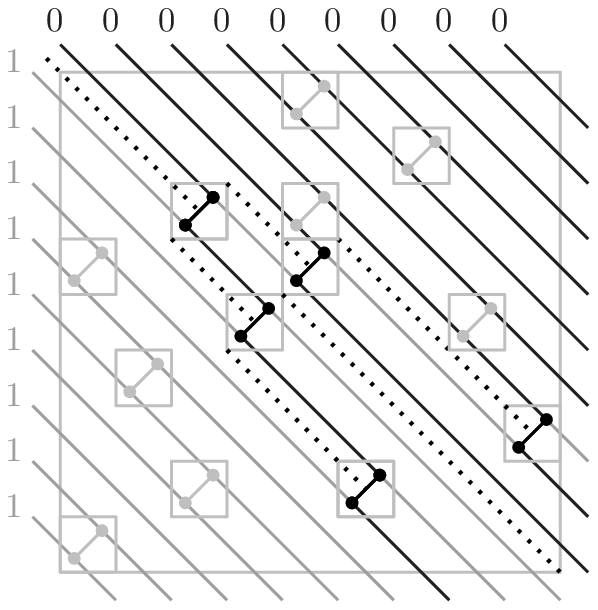}
\end{center}
  \caption{Comparing highly similar strings}
  \label{fig:highsim}
\end{minipage} 
\end{figure}
Figure~\ref{fig:zeroone_example} shows an example of the $\mathrm{LCSNET}(x,y)$ with 0/1 inputs
for the problem instance shown in Figure~\ref{fig:lcsexample} on page~\pageref{fig:lcsexample}.
It seems intuitive from this figure that the  stray zeros and ones trace contours in $L$.
\begin{theorem}\label{thm:straycellsvscontours}
A cell belongs to a contour in the matrix of prefix-prefix LCS lengths $L$ iff it has at least 
one stray value as an input or output.
\end{theorem}
\begin{proof}
This follows from Corollary~\ref{cor:0-1 net output} by induction on the number of contours. 
If $L$ has no contours, no match cells can exist. If there is exactly one contour in $L$, 
all match cells must belong to this contour, and the contour splits the set of cells
into two parts of mismatch cells. Consider the set of mismatch cells to the top/left of the contour.
All cells in this set have zeros as their top input and ones as their 
left input since these are either the input values to the transposition network, or
have been translated through the previous mismatch cells as shown in case (e) of 
Figure~\ref{fig:matchcategory}. All dominant matches on the contour must have a
zero as their top input and a one as their left input as well, since they must be at the right 
and below a case (e) mismatch cell, or equivalently at the top or left of the
alignment dag. Dominant match cells output a stray zero on the right and a stray one on the 
bottom (see case (d) in Figure~\ref{fig:matchcategory}). Any cell that has a stray zero
as its left input and a zero as its top input must be to the right of a match. As there is 
only one contour the cell cannot be below another match and therefore $L$ will increase vertically
in this cell since the prefix-prefix LCS can be extended by the first match to the left.
Symmetrically, this is true for any cell with a stray and a none-stray one as its inputs 
(see cases (a) and (b) in Figure~\ref{fig:matchcategory}). In the only remaining case, 
two stray values meet in the same cell $(\hat{\imath}, \hat{\jmath})$ (case (c) in 
Figure~\ref{fig:matchcategory}). In this case, the prefix-prefix LCS could either be extended 
by using the matches above $(\hat{\imath}, \hat{\jmath})$ or by using the matches to the left 
of $(\hat{\imath}, \hat{\jmath})$, but not by using both since they are incomparable under the
$\prec$ ordering (and no path containing one of each of those matches exists in 
the alignment dag). Now consider the cells immediately to the right or below the contour. 
These cells cannot be to the right or below dominant matches (otherwise they would belong
to the contour).
\begin{figure}[t]
\begin{center}
\includegraphics[width=0.85\textwidth]{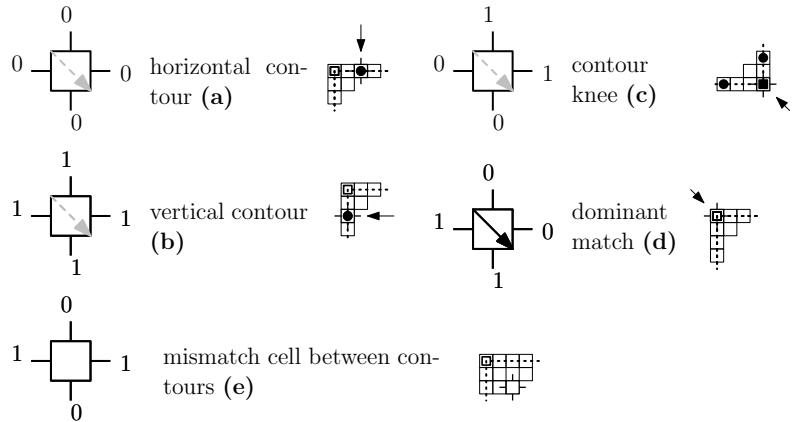}
\end{center}
\caption{Interpreting 0-1 transposition network cells and their inputs as contours.}\label{fig:matchcategory}
\end{figure}
Therefore, these cells must all have non-stray inputs (i.e.\ a zero at the top input and a one at the
left input), since cells on the horizontal contour output zeros on the bottom, cells on the 
vertical contour output ones at the right, and contour knees output a zero on the bottom 
and a one at the right output. As all the cells immediately neighbouring the contour to the 
right or below must be mismatch cells (only one contour exists $\Rightarrow$ all match cells 
are on it), they all belong to case (e) in Figure~\ref{fig:matchcategory} and in consequence
all cells below or to the right of them as well. Therefore, Theorem~\ref{thm:straycellsvscontours} 
is true in the case where only one contour exists. Furthermore, all additional contours
must either have case (e) cells on top and to their left, or border directly on another contour.
Cell contours output non-stray values on the right/the bottom if they have non-stray inputs.
Therefore, Theorem~\ref{thm:straycellsvscontours} is also true for more than one contour. 
$\Box$
\end{proof}
% Algorithm~\ref{alg:trans_antichains} shows how to compute the output of  
% $\mathrm{LCSNET}(x,y)$ with 0/1 inputs. 
The resulting algorithm is for computing output of  
$\mathrm{LCSNET}(x,y)$ with 0/1 inputs is equivalent to~\cite{Apostolico:87}, 
giving a running time of $O(mn\log p)$ which can be improved to
$O(m\log n + d\log (mn/d))$ using the finger searching technique~\cite{Brodal:05}. 

Consider the problem of comparing two strings that are highly similar.
Myers~\cite{Myers:86} proposed an algorithm to compare strings in time $O(ne)$,
where $e$ is the edit distance between the strings. The idea behind this algorithm
is to incrementally extend only the longest paths in the alignment dag until the
LCS is found. A similar algorithm can be obtained by using 0-1 transposition networks
as follows. 

If the two input strings are identical, no comparators exist on the main diagonal
of alignment dag cells, i.e.\ between transposition network wires $m$ and $m+1$.
This means that no ones can get to the right hand side, and no zeros can get to
the bottom of the alignment dag. We can look at this as two streams of zeros and
ones, and do not need to evaluate comparisons within a single stream of zeros or
ones. The only comparators which can possibly swap inputs are the ones between
streams. If a comparator occurs between two streams, the inputs will only be
swapped if the zero is input from the top, i.e.\ we can restrict our attention
to the upper boundaries of streams of ones. Figure~\ref{fig:highsim} shows an
example. The comparators drawn in black are those between streams of zeros and 
ones which must swap their inputs.
\begin{definition}
Let a \textit{1-0 boundary} in stage $s$ of $\mathrm{LCSNET}(x,y)$ with 0/1 inputs
be defined as any location in this stage where two adjacent wires $l$ and 
$l+1$ carry values one and zero respectively.
\end{definition}
\begin{corollary} The number of 1-0 boundaries in any stage of the transposition
network is dominated by $k + 1 = n - p + 1$.\label{cor:1-0boundaries}
\end{corollary}  
\begin{proof}
By induction: Assume $m = n = 1$. The transposition network has two wires which are
initialized with a zero and a one. Therefore, the number of 1-0 boundaries must be
less or equal than $1$. The LCS distance $k$ can be $0$ or $1$.
Increasing $m$ or $n$ by one adds another row or column of comparators to 
the transposition network. 
Consider the case of adding a column of comparators (i.e.\ increasing $n$ by one). 
Each $1$ which is output at the right hand side can only cause one 1-0 boundary. 
Furthermore, ones do not move downwards in comparisons. Therefore, a new 1-0 boundary 
can only be created if a value of $1$ from the left hand side reaches the right hand side, 
which means that the number of 1-0 boundaries cannot increase by one in this case 
without also increasing $k$ by one. However, $k$ cannot increase by more than one,
since maximally a single value of $1$ reaches the right hand side. Symmetrically,
when increasing $m$ by one, we add a row of comparators at the bottom. If we have
$k$ zeros at the bottom, each of these zeros can only be part of a single 1-0 boundary.
We can only gain a single $0$ on the bottom by increasing $m$ by one, in which case
also $k$ increases. Therefore $k+1$ always dominates the number of 1-0 boundaries.
$\Box$
\end{proof}

Using this insight, the LLCS of two strings $x$ and $y$ with $|x| = |y| = n$ can be 
computed in time $O(nk)$. 
This is done by tracing the intersections of the 1-0 boundaries with the 
$2n-1$ antidiagonals of the alignment dag, as this is the only place where change can 
occur. By Corollary~\ref{cor:1-0boundaries}, we know a bound for the number of 1-0 boundaries.
At each intersection of a 1-0 boundary with an antidiagonal, the corresponding characters
in $x$ and $y$ must be compared to check whether a comparator exists. This can be done in 
constant time, and since there are $2n-1$ antidiagonals we get the claimed running time. Note
that this algorithm does not require any pre-processing to obtain match lists. 

\begin{corollary}\label{cor:dominant on 0-1 boundary}
All dominant matches must be on a 1-0 boundary in the transposition
network.
\end{corollary}
\begin{proof}This follows immediately from Theorem~\ref{thm:straycellsvscontours}. 
\end{proof}
Corollary~\ref{cor:dominant on 0-1 boundary} allows to narrow down the area in
which to search for dominant matches, and can be used to extend Algorithm~\cite{Apostolico:87} 
to achieve running time $O(kp)$, similarly to~\cite{Rick:95}.

\begin{theorem}
The implicit highest-score matrix for comparing two strings of length $n$ can be computed in
time $O(np)$.
\end{theorem}
\begin{proof}
Using the 0-1 transposition network, we are able to determine for every match
cell whether it is dominant or non-dominant, as well as for every mismatch cell whether it 
is part of a contour.
Looking at this in the more general setting of semi-local string comparison where we need to trace all
seaweeds individually, we can still see that non-trivial comparisons between
seaweeds can only occur when the cell is actually part of a contour.
Cells outside the contours are always mismatch cells which compare an input originating at the
left hand side of the alignment dag to an input originating at the top of the alignment dag. Therefore all the
comparators in these cells can be replaced by swap operations (i.e.\ they contain seaweed crossings). 

Given all dominant matches on a contour and the values on all transposition network wires before
they intersect the contour, we can compute the values on all wires of transposition network after 
the intersection in time which is linear in the length of the contour. As all comparators
between contours perform swap operations, we can also compute the permutation of values
performed between two contours in time linear in the length of the longer contour.  

It is possible to compute the set of all $k$-dominant matches with $k \in [1:p]$ in $O(np)$ time. 
We can use Algorithm~\cite{Apostolico:87}\footnote{A practical algorithm for computing a 
list of dominant matches is described in~\cite{Crochemore+:03}} for this. Knowing the dominant matches 
in every antichain, we can trace its complete contour in time linear in its length. No contour can 
have length $l$ longer than $2n$, and there are exactly $p = \mathrm{LLCS}(x,y)$ contours. 
Further, we can obtain the inputs and outputs of all cells in a contour of length $l$ in time $O(l)$
with $l \leq 2n$. Therefore, the worst case running time of our algorithm for semi-local string comparison 
is bounded  by $O(np)$. $\Box$ 
\end{proof}

\section{Conclusions}
In this paper, we have presented a new method of solving the semi-local
string comparison problem using transposition networks. This method provides a
unified view of different string comparison algorithms, and allows to obtain
efficient algorithms for global string comparison which have the same
complexity as the best known algorithms.
Furthermore, we have obtained new algorithms for sparse semi-local string
comparison, high similarity and dissimilarity string comparison, as well as
semi-local comparison of run-length compressed strings. 
In a separate paper, we will show that it is possible to implement the algorithms 
for semi-local string comparison efficiently for an application
to LCS-filtered dot-plots~\cite{Maizel_Lenk:81}.
We conclude that the transposition network method is a very general and flexible way of
understanding and improving different string comparison algorithms.

\bibliographystyle{splncs}

\end{document}